# Effect of Resonant Acoustic Powder Mixing on Delay Time of W-KClO$_4$-BaCrO$_4$ Mixtures

*Kyungmin Kwon [a, ‡], Seunghwan Ryu [a, ‡], Soyun Joo [a, ‡], Youngjoon Han [a], Donghyeon Baek [b], Moonsoo Park[b], Dongwon Kim[b] and Seungbum Hong [a, *]*

[a] Department of Materials Science and Engineering, KAIST, Daejeon 34141, Republic of Korea

[b] Poongsan Defense R&D Institute, Gyeongju-si, 38026, Republic of Korea

[‡]These authors contributed equally to this work.

*Email address: seungbum@kaist.ac.kr

ABSTRACT

This study investigates the impact of resonant acoustic powder mixing on the delay time of the W-$KClO_4$-$BaCrO_4$ (WKB) mixture and its potential implications for powder and material synthesis. Through thermal analysis, an inverse linear relationship was found between thermal conductivity and delay time, allowing us to use thermal conductivity as a reliable proxy for the delay time. By comparing the thermal conductivity of WKB mixtures mixed manually and using acoustic powder mixer, we found that acoustic powder mixing resulted in minimal deviations in thermal conductivity, proving more uniform mixing. Furthermore, DSC analysis and Sestak-Berggren modeling demonstrated consistent reaction dynamics with a constant activation energy as the reaction progressed in samples mixed using acoustic waves. These findings underscore the critical role of uniform powder mixing in enhancing the thermodynamic quality of the WKB mixture and emphasize the importance of developing novel methods for powder and material synthesis.

1. **Introduction**

Ammunition serves as a crucial element within weapons, capable of directly dismantling materials and facilities, thereby significantly influencing the outcome of warfare [1]. Recent events, such as the conflict in Ukraine, have underscored the heightened significance of ammunition. Meanwhile, assessing the stability and performance of ammunition presents challenges, as it is typically disposed of after a single use, rendering direct examination impractical and unfeasible. Consequently, the ammunition design process necessitates a multidisciplinary approach that incorporates chemical analysis, modeling, and DFT calculation to ensure optimum outcomes [2].

The structure of ammunition consists of various components, including a detonator, propellant, projectile, main charge, fuse, among others, with their placement and chemical composition varying based on the type. Typically, ammunition employs specific chemicals, such as a detonator gunpowder that is ignited upon impact to generate initial energy, an ignition agent to ignite a delay mixture using the energy of the primer gunpowder, and a delay mixture that induces controlled delay in the event of combustion. These components are loaded within the bullet body in the sequence of detonator gunpowder, ignition agent, and delay mixture, where energy is transmitted in that order to fulfill the function of ammunition [3]. Among these components, the delay mixture serves as a chemical timer to ensure a safe distance between the muzzle and ignition by introducing a time delay effect. Consequently, the delay time is a vital quality indicator that ammunition users and producers must consider, and demands consistency and uniformity in its implementation.

A typical delay mixture is comprised of solid powders consisting of a fuel and an oxidizer. Its primary objective is to initiate a chain explosion by providing energy that exceeds the activation energy required by the detonator through an exothermic reaction. To accomplish this, the delay mixture incorporates an oxidation-reduction reaction, which releases substantial energy as an exothermic process. Furthermore, as ammunition often undergoes extended periods of storage, it is imperative to mitigate the impact of external environmental factors such as temperature and humidity on the delay mixtures to ensure sustained performance over time.

Traditionally, delay mixtures have been categorized into two types based on gas generation: those that produce gas and those that do not. Delay mixtures that generate gas are employed in inexpensive ammunitions such as signal flares, where precise control of the reaction rate is not crucial. Conversely, other types of ammunition utilize gas-free delay mixtures to prevent the alteration of combustion rates caused by changes in internal pressure of the ammunition due to gas generation. Types of commercially available delay mixtures include tungsten-based, zirconium-nickel-based, and chromium-based variants, among others. Among these, the WKB delay mixture consisting of tungsten (W), potassium perchlorate ($KClO_4$), and barium chromate ($BaCrO_4$) exhibits physicochemical properties favorable for use as a delay mixture in a wide range of mixing ratios. A significant advantage of the WKB delay mixture lies in its ability to achieve desired delay times by adjusting the combustion characteristics based on particle size. Consequently, many research groups [4 - 10] have investigated the combustion mechanism of the WKB delay mixture through thermal analysis techniques. These investigations revealed that the primary reaction involves the oxidation of W by $KClO_4$. In addition, Nakamura et al. demonstrated that decreasing the W content in the tungsten-based delay mixture increases the delay time, establishing a correlation between the thermal conductivity of the delay mixture and the delay time. Moreover, the respective contributions of each component to the delay time was determined to be in the order of significance: $BaCrO_4$ > $SiO_2$ > $KClO_4$ [6].

However, tungsten-based delay mixtures exhibit significant variations in delay time depending on the aged state of its constituent raw materials, leading to challenges in securing uniform quality [19, 20]. Therefore, the storage and management of raw materials have emerged as crucial considerations to ensure consistent delay time quality for tungsten-based delay mixtures. As such, precise control over the process and environmental factors becomes vital for the purpose of controlling the quality of tungsten-based delay mixtures. Moreover, given that the delay time of the mixture is primarily determined during the manufacturing process, meticulous control of the particle size and distribution of each component is of utmost importance. In pursuit of this objective, we delved into the utilization of resonant acoustic waves as a means to augment mixing efficiency. Multiple studies have illustrated the efficacy of resonant acoustic mixing (RAM) in enhancing the homogeneity of powder mixing procedures, consequently bolstering both the mechanical and chemical stability of the resultant sample in various applications, including powder blends of pharmaceutical compounds [11], energetic materials [12], industrial materials [18], and the scale-up production of cocrystals [13].

In this study, we followed the conventional industrial procedure to prepare WKB specimens, both mixed by hand and mixed using a RAM method. We conducted thermal and physical property analyses, including measurements of thermal conductivity, true density, and analysis of thermal properties to determine the reaction activation energy. Our studies reveal that when utilizing resonant acoustic waves for mixing, uniform reaction dynamics occur throughout the reaction, accompanied by a constant activation energy. This observation highlights the critical influence of uniform powder mixing on the quality of the WKB mixture. Consequently, our findings strongly suggest the necessity to enhance the powder mixing method as a means to improve overall quality.

## 2. Experimental details

2.1. Sample preparation

Solid samples with a diameter of 5 cm and a thickness of 1.3 cm were prepared by compressing W, KClO$_4$, BaCrO$_4$, and binder powder. W, BaCrO$_4$, and KClO$_4$ served as the main raw materials, while diatomaceous earth and binder were used as additives. Each raw material was weighed and placed in a designated container, followed by the addition of the binder solution. Conventionally prepared samples, referred to as 'Reference' samples, was thoroughly wet mixed by hands covered with nitrile gloves, followed by a drying and sieving process.

To compare the results, powder mixing using acoustic waves was conducted using acoustic resonant waves (Resodyn LAMBRAM II, Resodyn Acoustic Mixers) and underwent a similar manufacturing process. These specimens are referred to as 'Acoustic' samples. RAM mixers operate at a nominal frequency of 60 Hz, which can vary slightly depending on the resonance frequency of the whole system, including the product being mixed. The amplitude of the movement is 0.500 (1.27 cm) or less and the acceleration was selected to be 80 g. The rationale for opting for a relatively substantial acceleration lies in the potential amplification of RAM performance with increasing acceleration levels. [14].

In our investigation, it was confirmed that the implemented processes in the case of Reference samples did not introduce any external components, as no traces of nitrile gloves used during handling were detected.

2.2. Thermal analysis

Thermal properties, including thermal conductivity, thermal diffusivity, specific heat, and thermal absorption of all of the samples were measured using the hot-disk method (TPS 3500, Hot Disk AB, Göteborg, Sweden). All Differential Scanning Calorimeter (DSC) curves were measured using a DSC 404 F1 from NETZSCH. Measurements were performed in Ar to prevent nitride formation. The temperature range was from 30 °C to 800 °C, with the heating rate varied from 2, 5, 10, 15, to 20 °C per minute.

2.3. Delay time measurement

The delay tube is charged with a limited amount of initiator and a specific amount of delay mixture, with added igniter materials. To initiate the process, a detonator is used to trigger the igniter positioned on top of the delay tube. The igniter begins to ignite, driven by the energy generated by the operating detonator. Once ignited, the delay mixture burns gradually, ensuring the required delay time to be achieved.

Here, the delay time refers to the duration between the ignition of the igniter and the completion of the delayed combustion. To measure the delay time accurately, we relied on two key factors: the sound produced by the initiator operation when the igniter is fired, and the time it takes for the delayed combustion to reach its conclusion.

2.4. Kinetics of reaction

In this study, in order to gain a deeper understanding of the reaction mechanism involved in complex reactions, we utilized the kinetic parameters derived from the multi temperature-rate thermogravimetric analysis and differential scanning calorimetry (TG-DSC) measurements. An isoconversion technique was employed to express the activation energy of each reaction as a function of reaction progress, according to the methodology described by Han et al. [9]. Briefly, this method allows the effective activation energy (Ea) to be measured without the need to select a specific reaction model, where variation in Ea can generally be associated with a change in the reaction mechanism or the rate-limiting step of the overall reaction [15]. Our calculative methods are based on Equation (1), which is derived by application of the Arrhenius equation that describes the kinetics of the solid-phase reaction, followed by appropriate substitutions of the relevant thermodynamic factors.

$$ln \ln \left(\frac{d\alpha}{dt}\right) = ln \ln(A) + m \cdot ln \ln(\alpha) + n \cdot ln \ln(1-\alpha) - \frac{E_\alpha}{RT} \quad\quad Equation \quad (1)$$

where α is reaction progress, t is time, A is pre-exponential factor, T is absolute temperature, m is the factor contributing to the nucleation dominant process, and n is the reaction order of the chemical process. Details regarding this equation are further provided in the previously mentioned work [9].

## 3. Results and discussion

As aforementioned, the WKB-based delay tube comprises an ignition agent, an initiator, and primary, secondary, and tertiary delay mixtures, where the delay time depends on the specific set of WKB specimens used for each delay mixture. In particular, considering the linear proportionality between the burning rate and the W content [2, **Figure. S1**], we can deduce an inverse relationship between the delay time and the amount of W present. Furthermore, based on the positive correlation between logarithm of thermal diffusivity and the logarithm of burning rate, [**Figure. S2**], it becomes evident that the delay time decreases as the thermal diffusivity increases.

To establish a suitable proxy to represent the delay time, this study involved the measurement of thermodynamic factors such as thermal conductivity, thermal diffusivity, specific heat, and heat absorption for each sample. Among these factors, thermal conductivity exhibited the highest reliability in terms of repeatability (**Figure. S3**). Thus, it was selected as a proxy to represent the delay time. This relationship between these two parameters (thermal conductivity and delay time) has also been dealt with in [16]. Additionally, since thermal conductivity is linearly proportional to thermal diffusivity, we extend the trend reported in the literature to conclude that the delay time is inversely proportional to thermal conductivity.

**Figure. 1** shows the results of delay time measurements for three types of delay tubes manufactured using different W types and weight fractions (wt%). The composition of each delay tube is provided in **Table 1**, where the charging ratio means the fraction of each composition loaded into the tubes and the primary mixture represents the one that is ignited first.

**Table 1.** The percentage of tungsten delay mixtures for each delay type

|  | Delay mixture |  |  |  |  |  |
|---|---|---|---|---|---|---|
|  | **Primary delay mixture** | | **Secondary delay mixture** | | **Tertiary delay mixture** | |
|  | Charging Ratio | W wt% | Charging Ratio | W wt% | Charging Ratio | W wt% |
| **Delay Tube 1** | 70 | 28 | 15 | 31 | 15 | 40 |
| ***Delay Tube 2** | 70 | 28 | 15 | 31 | 15 | 40 |
| **Delay Tube 3** | 70 | 30 | 30 | 35 | - | - |

*The type of W is different from that of Delay Tube 1.

By performing linear fitting based on the measured thermal conductivity and delay time, we confirmed the inverse proportional relationship between the two variables, as discussed earlier ($R^2 = 0.71$). Notably, this tendency holds true for all delay tubes, even in the case of delay tube 3, which consists only of the primary and secondary delay mixtures without a tertiary delay mixture. This suggests that the relationship between thermal conductivity and delay time is not solely dependent on the specific elements used in the sample, but rather on the thermal conductivity of the entire sample comprising the delay time. In addition, although delay tube 3 exhibits a slightly higher average W content compared to delay tubes 1 and 2, the reason for its longer delay time and lower thermal conductivity lies not only in the W composition, but also in other factors such as particle size uniformity (the results of these tests are not shared within the scope of this paper). This implies that additional factors beyond the W content contribute to determining the delay time of a certain tube.

However, measuring the delay time for each experiment and set of specimens is a laborious and resource-intensive process, requiring multiple and repeated measurements for reliable data. Consequently, our established relationship between thermal conductivity and delay time is promising in that it eliminates the need for repetitive delay time measurements. Instead, by simply measuring thermal conductivity, the delay time can be predicted. This development significantly reduces the hassle associated with obtaining delay time and streamlines the evaluation of various approaches for controlling delay times in the future. Moreover, this simplified method of assessing delay tube quality based on thermal conductivity enables easier verification of the effectiveness of different techniques or processing steps applied for delay time manipulation and product quality control.

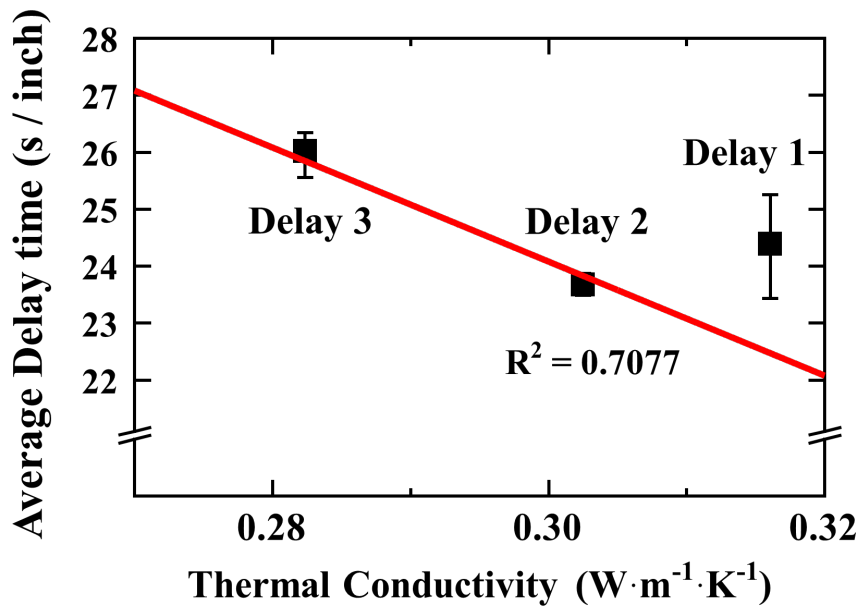

**Figure. 1.** Relationship between thermal conductivity of WKB mixture and delay time.

The control of delay time in the delay mixture requires analysis of the combustion reaction, as it directly influences the delay time. Several indicators, including thermal conductivity, thermal diffusivity, specific heat, and heat absorption rate, are expected to play a crucial role in the combustion reaction, particularly in heat transfer processes [17]. These are also thermal properties that can be obtained by the hot-disk method.

**Figure. 2** presents the measurements of these indicators for various samples with different types and fractions of tungsten, and the relationship between these indicators are summarized in equations in **Equations (2) – (5)** (the raw data of these measurements are shown in **Figure. S3**). Previous studies have established the inverse relationship between thermal diffusivity and delay time [4]. However, our measurements revealed significant dispersion in thermal diffusivity even within the same sample. Meanwhile, examination of **Equation 3** shows that thermal diffusivity is determined by thermal conductivity (k), specific heat ($c_p$), and true density (ϱ). Upon reviewing the measurement results, we observed that only specific heat exhibited significant dispersion among the three indicators. A trend similar to that of specific heat was derived, suggesting that the variation in specific heat affects the overall measurement of thermal properties with the exception of thermal conductivity as cp can be found as a key parameter to calculate other thermal indicators as seen in **Equations (2) – (5).** This dispersion in specific heat can be attributed to the differences in the amount of heat absorbed during the temperature rise process, indicating a non-uniform combustion reaction.

In contrast, thermal conductivity exhibited minimal dispersion within each sample compared to other thermal parameters, demonstrating high reliability and repeatability. The thermal conductivity of a sample is determined by the shortest heat conduction path within the material, which tends to converge to a consistent value (i.e. mean distance) regardless of dynamic factors such as powder distribution uniformity. Consequently, constant thermal conductivity measurements can be obtained for varied trials of the same sample. Due to its suitability as an index for explaining our combustion reaction model, thermal conductivity was adopted and used as the representative indicator in this paper.

Since tungsten has higher thermal conductivity compared to the oxidizing agent or binder materials present, the expectation was that thermal conductivity would increase with higher fractions of W. However, measuring the thermal conductivity of the samples that did not undergo acoustic processing revealed that the tertiary delay mixture (W 40 wt%) had lower thermal conductivity than the second delay mixture (W 31 wt%). (**Figure. 2.a**, reference samples indicated as gray points). This result contradicts the initial hypothesis.

On the other hand, when samples were subjected to acoustic powder mixing, a clean linear relationship was observed between W fraction and thermal conductivity (**Figure. 2.a**, acoustic powder mixed samples indicated as red points). This can be attributed to the uniformity of heat conduction paths achieved through even mixing of the powder sample and the consequent stable heat conduction process. It is inferred that by adjusting the weight fraction of W, it would even be viable to produce a sample with desired thermal conductivity. This finding is expected to enable effective control over the delay time of the WKB mixture for industrial purposes.

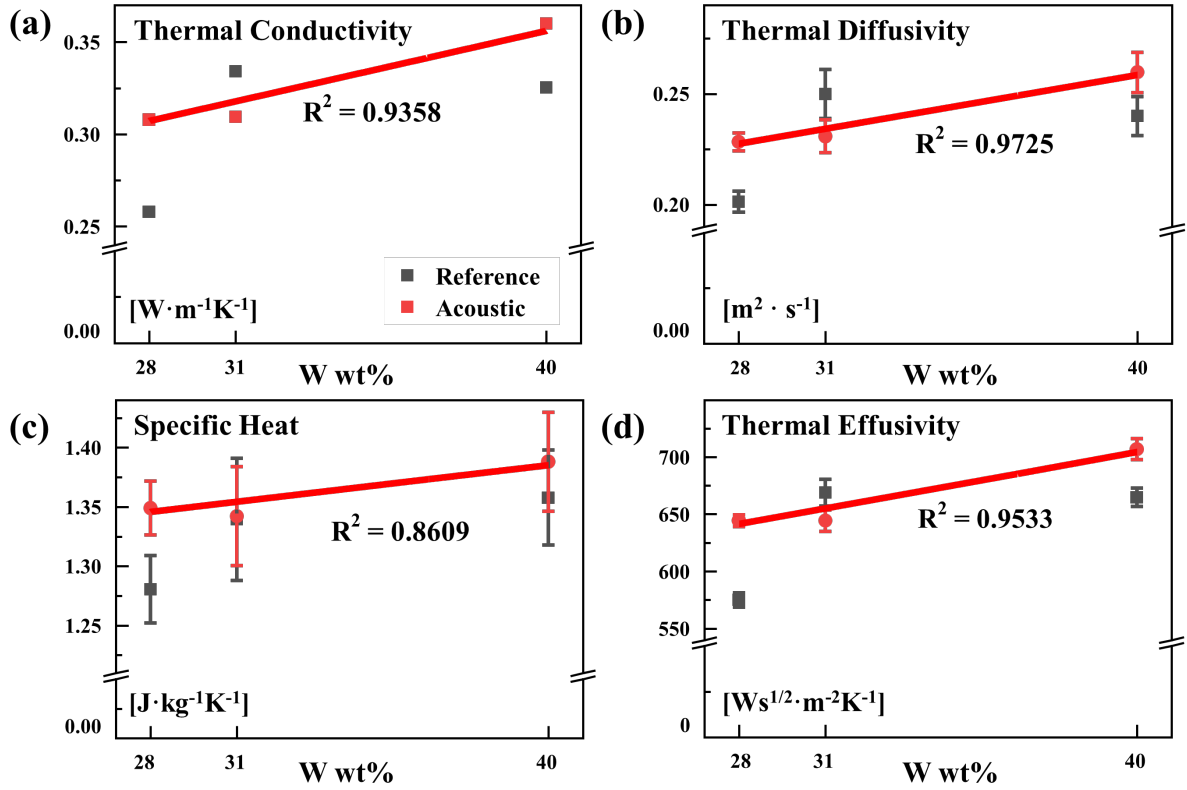

**Figure. 2.** Thermal properties of WKB mixtures measured by a hot disk thermal conductivity meter. Red line represents the linear fit of each thermal property for the acoustic powder mixing processed sample.

$$\text{Thermal conductivity}, k = \frac{PL}{A\Delta T} \qquad Equation \quad (2)$$

$$\text{Thermal diffusivity}, \alpha = \frac{k}{\rho c_p} \qquad Equation \quad (3)$$

$$\text{Specific heat}, c_p = \frac{Q}{m\Delta T} \qquad Equation \quad (4)$$

$$\text{Thermal effusivity}, e = \sqrt{k\rho c_p} \qquad Equation \quad (5)$$

As mentioned earlier, the thermal conductivity measurement indicated a more uniform distribution with acoustic powder mixing. To confirm that the chemical reaction also proceeds uniformly, we additionally measured the change in activation energy according to the reaction process. The thermal properties of the WKB mixture were analyzed using DSC, following the specific theories and methods outlined in **Part 2.4** [9].

DSC analysis was performed in Ar in the temperature range of 30 °C to 800 °C with five different heating rates. **Figure. 3a** shows the DSC curve obtained by gradually increasing the temperature of the 1R sample (primary delay mixture in delay tube 1 with acoustic mixing, W 28 wt%) at a rate of 10 ℃ per minute. The DSC curves of the 1R sample at other heating rates are shown in **Figure. S4**. Within the elevated temperature range, one endothermic peak, one largest exothermic peak, and two small exothermic peaks were observed. It is known that the endothermic peak observed around 304 ℃ is caused by the crystal phase transition process of the oxidizing agent $KClO_4$ from a rhombic structure to a cubic structure. The peaks occur between 445℃ and 485℃, and between 465℃ and 514℃, depending on the heating rate, overlapped with the decomposition of $KClO_4$ as well as the main combustion reaction, which is the combustion reaction between tungsten and oxygen (**Figure. S4**). The exothermic peaks occurring at the temperature after the main combustion reaction are due to the decomposition and oxidation reactions of the remaining, unreacted oxidizing agent [10].

Regarding the case of the reaction progress $\alpha$, values increased by 0.1 units from 0.1 to 0.9 were selected [9]. By extending the logarithmic reaction rate plot, we calculated the corresponding slope and y-intercept. The activation energy was obtained by multiplying the obtained slope by the constant R. We expressed this activation energy as a function of the reaction progress $\alpha$, and obtained graphs as shown in **Figure. 3b – c** to assess the uniformity of the reaction kinetics throughout the reaction process. These sets of graphs demonstrate the power generation behavior in accordance with the progress of the combustion reaction. The detailed schematic of the surface reaction according to reaction progress are provided in **Figure. S5**.

Furthermore, we plotted the change in activation energy and the derivative of the activation energy with respect to the reaction process with the delay mixtures containing 28 wt% and 31 wt% of tungsten. In a uniform chemical reaction, both the activation energy and the its derivative should remain constant over the course of the chemical reaction. Without acoustic treatment, activation energy showed fluctuations, either increasing or decreasing as the chemical reaction progressed (**Figure. 3b**). However, in case of acoustic powder mixing processed sample, the activation energy either decreased or remained relatively constant (**Figure. 3c**). In terms of the first derivative of the activation energy, the reference sample exhibited a consistent transition from a positive to a negative value, whereas the processed sample converged to zero and remained constant (**Figure. 3d-e**). These observations demonstrate that homogeneous mixing contributes to the stabilization and homogenization of multi-state reactions, facilitating control and regulation of combustion reactions.

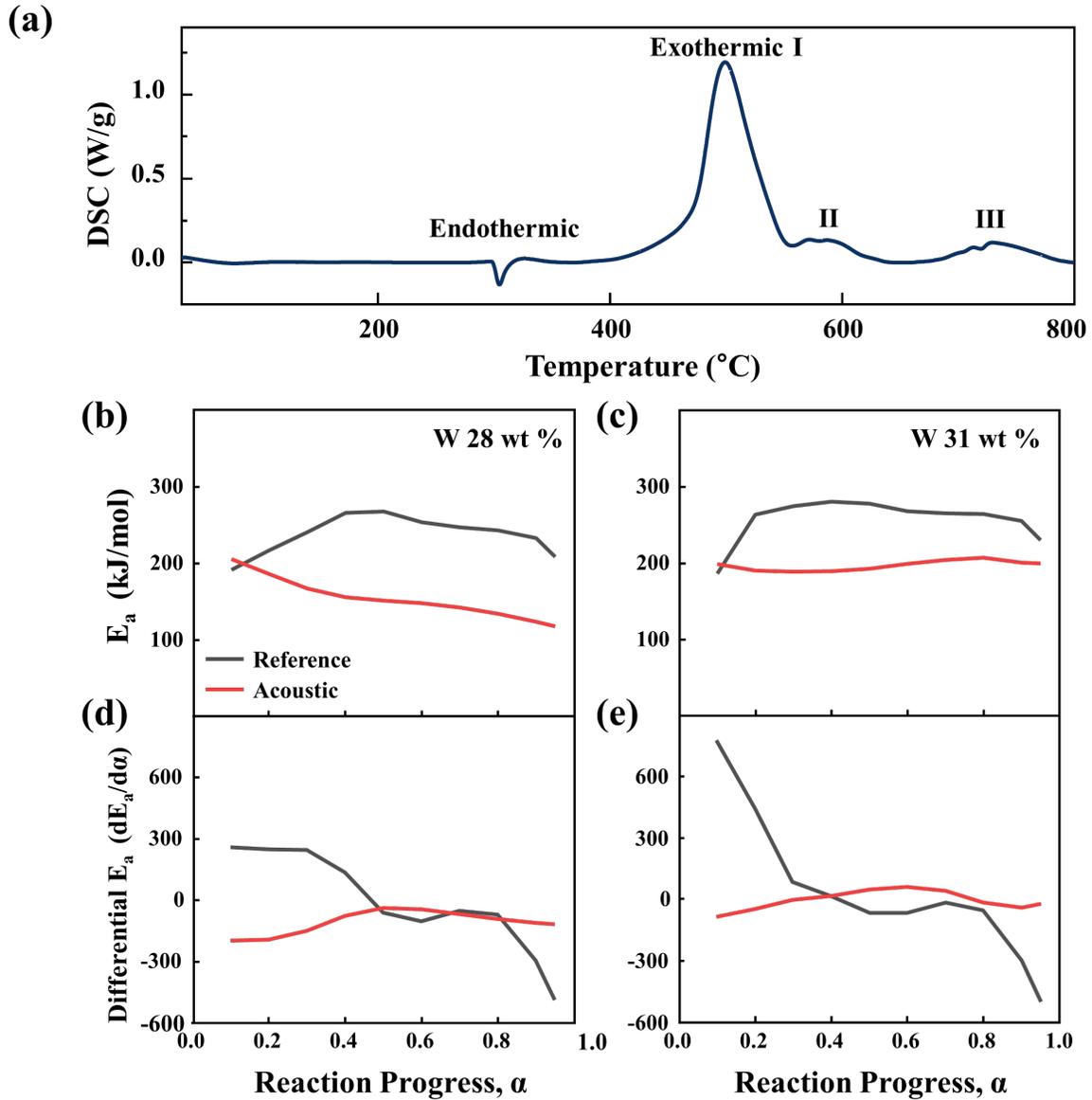

**Figure. 3**. (a) Differential scanning calorimetry (DSC) curves of WKB mixture sample 1R with heating rate 10 °C/min, (b–c) Activation energy ($E_a$) vs. reaction progress, (α) for main combustion peaks of WKB mixtures W 28 wt% and 31 wt% (Reference vs. Acoustic Samples), (d–e) Differential $E_a$ vs. reaction progress of (b-c)

4. **Conclusions**

The present study investigated the impact of acoustic mixing on the delay time of the WKB mixture through thermal characterization. The findings clearly indicate that thermal conductivity serves as an important indicator for predicting delay time. Moreover, when employing acoustic waves for mixing, the deviation in thermal conductivity was minimal, aligning closely with the predicted values based on W fraction. Additionally, thermal property analysis using DSC revealed a more uniform progression of the chemical reaction with respect to the reaction progress. These results underscore the critical role of the mixing method in controlling the quality of WKB delay mixtures and offer promising prospects for the advancement of high-quality delay mixtures in future applications.

AUTHOR INFORMATION

**Corresponding Author**

Seungbum Hong – Department of materials Science and Engineering, Korea Advanced Institute of Science and Technology, Daejeon 34141, Republic of Korea; orcid.org/0000-0002-2667-1983; Email: seungbum@kaist.ac.kr

**Author Contributions**

The manuscript was written through contributions of all authors. All authors have given approval to the final version of the manuscript. ‡These authors contributed equally.

**Notes**

The authors declare that they have no known competing financial interests or personal relationships that could have appeared to influence the work reported in this paper.

ACKNOWLEDGMENT

This research was supported by "Poongsan-KAIST Future Research Center Project" and the authors gratefully acknowledge the financial support of Poongsan Corporation. This work is in part supported by the National Research Foundation of Korea (NRF) Grant, funded by the Korean Government (NRF-2022K1A4A7A04095890), for supporting the Korean-German collaboration. The authors are also grateful to KAIST Analysis Center for Research Advancement (KARA) for conductivity meter and DSC analysis.

# Supporting Information for

# Effect of Resonant Acoustic Powder Mixing on Delay Time of W-KClO$_4$-BaCrO$_4$ Mixtures

*Kyungmin Kwon [a,1], Seunghwan Ryu [a,1], Soyun Joo [a,1], Youngjoon Han [a],*

*Donghyeon Baek [b], Moonsoo Park[b], Dongwon Kim[b] and Seungbum Hong [a, *]*

[a] Department of Materials Science and Engineering, KAIST, Daejeon 34141, Republic of Korea

[b] Poongsan Defense R&D Institute, Gyeongju-si, 38026, Republic of Korea

[1]These authors contributed equally to this work.

*Email address: seungbum@kaist.ac.kr

**Content:**

Figure S1: Relationship between the burning rate and the tungsten content for a W/KClO$_4$/BaCrO$_4$ delay composition. Reproduced with permission from [4]. Copyright Korean Society of Explosives and Blasting Engineering (2000)

Figure. S2. Relationship between burning rate and thermal diffusivity of the W-KClO$_4$-BaCrO$_4$ delay combustion. Reproduced with permission from [4]. Copyright Korean Society of Explosives and Blasting Engineering (2000).

Figure. S3. Thermal properties of WKB mixtures measured by hot disk thermal conductivity meter.

Figure. S4. Multiple heating rates DSC curves of WKB mixture (2, 5, 10, 15, 20 °C / min).

Figure. S5. Schematic of surface reaction and activation energy change according to reaction progress

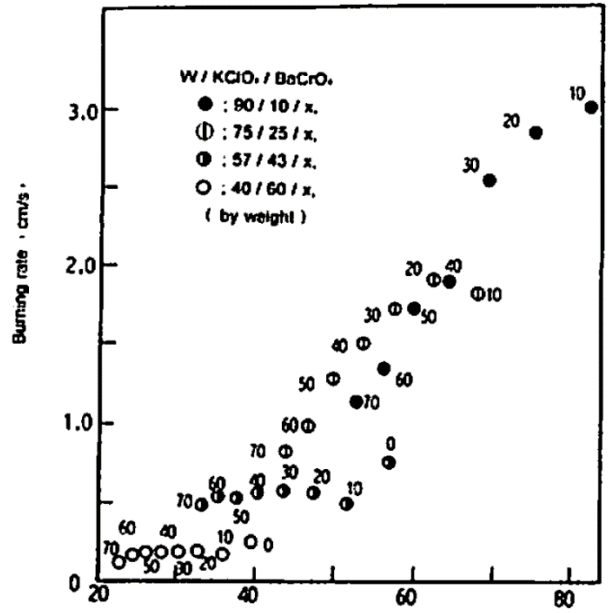

Figure. S1. Relationship between the burning rate and the tungsten content for a W/KClO$_4$/BaCrO$_4$ delay composition. Reproduced with permission from [4]. Copyright Korean Society of Explosives and Blasting Engineering (2000).

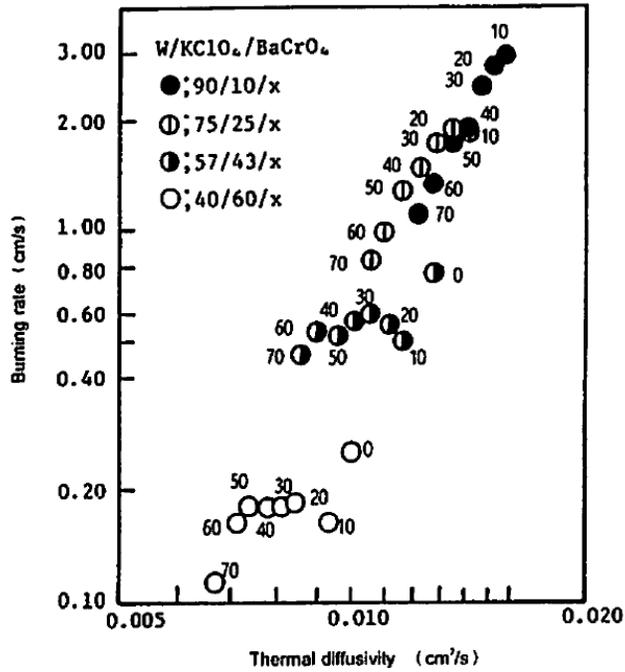

Figure. S2. Relationship between burning rate and thermal diffusivity of the W-KClO$_4$-BaCrO$_4$ delay combustion. Reproduced with permission from [4]. Copyright Korean Society of Explosives and Blasting Engineering (2000).

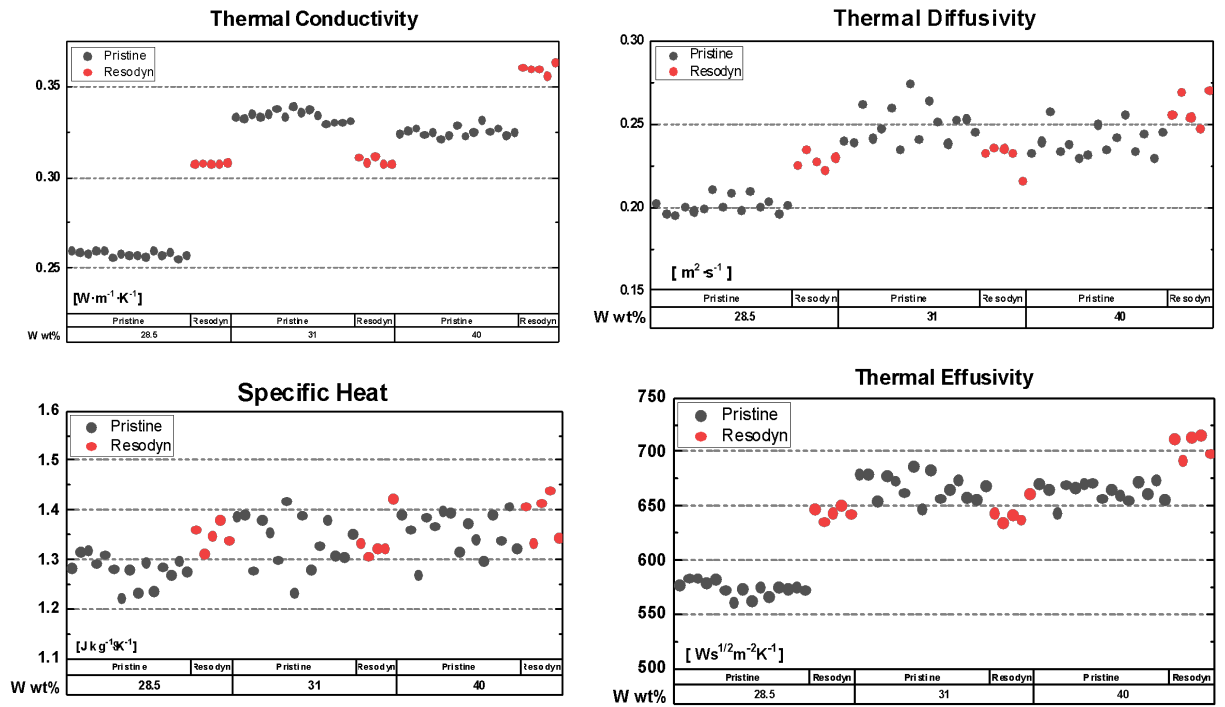

Figure. S3. Thermal properties of WKB mixtures measured by hot disk thermal conductivity meter.

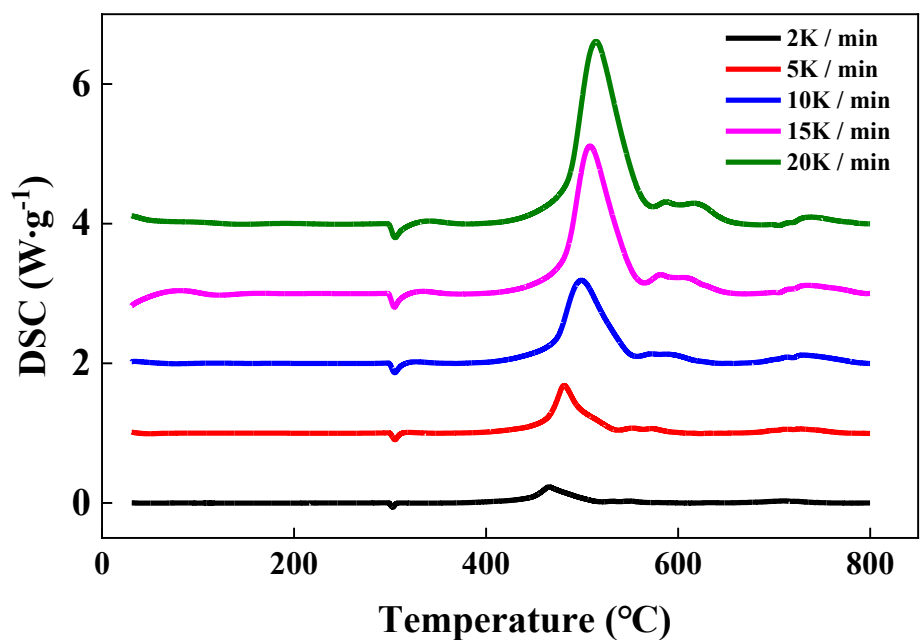

Figure. S4. Multiple heating rates DSC curves of WKB mixture (2, 5, 10, 15, 20 °C / min).

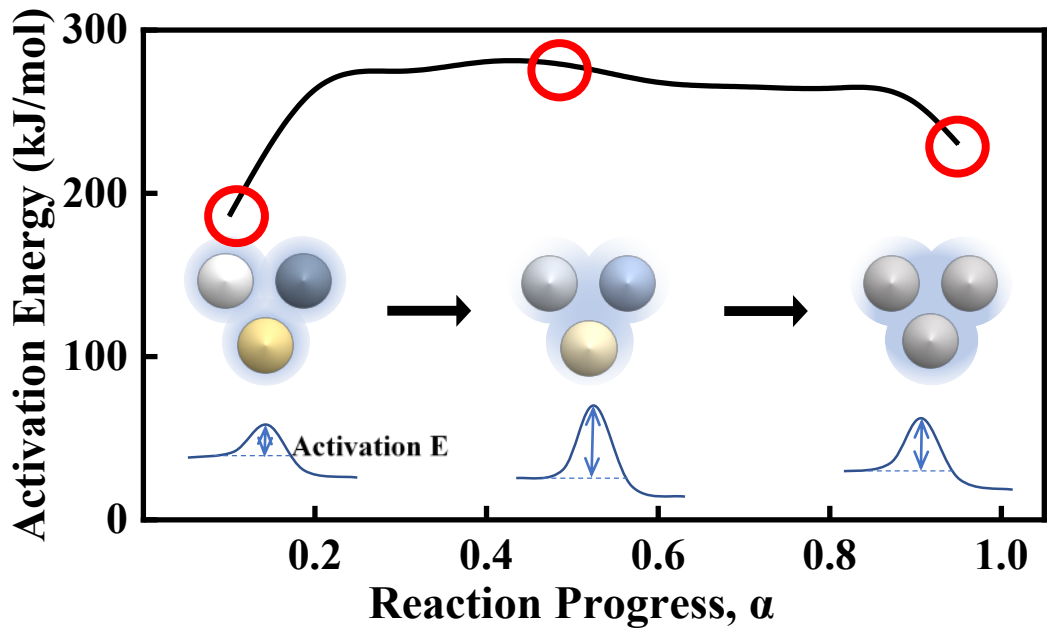

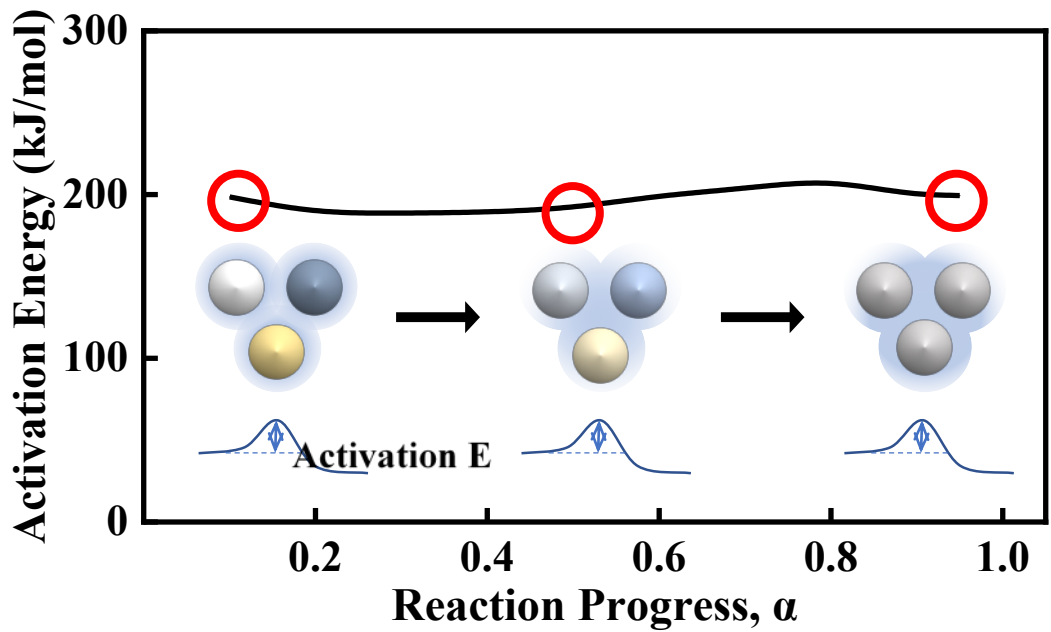

Figure. S5. Schematic of surface reaction and activation energy change according to reaction progress